\documentclass[aps,prl,reprint,superscriptaddress,amsmath,amssymb,floatfix]{revtex4-2}

\usepackage{graphicx}
\usepackage{bm}
\usepackage{braket}
\usepackage[dvipsnames]{xcolor}
\usepackage{float}
\usepackage[
    colorlinks=true,
    linkcolor=blue,
    citecolor=blue,
    urlcolor=blue
]{hyperref}

\graphicspath{{figures/}}
\begin{document}

\title{Microscopic Realization of Topologically Quantized Alignment in Fast-Rotating Nuclei}

\author{Ganlong Ding}
\affiliation{School of Physics and Astronomy, Beijing Normal University, Beijing 100875, China}
\affiliation{Department of Physics, Graduate School of Science, The University of Tokyo, Tokyo 113-0033, Japan}

\author{Sibo Wang}
\affiliation{Department of Physics and Chongqing Key Laboratory for Strongly Coupled Physics, Chongqing University, Chongqing 401331, China}
\affiliation{Department of Physics, Graduate School of Science, The University of Tokyo, Tokyo 113-0033, Japan}

\author{Hiroyuki Tajima}
\affiliation{Department of Physics, Graduate School of Science, The University of Tokyo, Tokyo 113-0033, Japan}
\affiliation{Quark Nuclear Science Institute, The University of Tokyo, Tokyo 113-0033, Japan}
\affiliation{RIKEN Nishina Center, Wako 351-0198, Japan}

\author{Daisuke Suzuki}
\affiliation{Department of Physics, Graduate School of Science, The University of Tokyo, Tokyo 113-0033, Japan}
\affiliation{Quark Nuclear Science Institute, The University of Tokyo, Tokyo 113-0033, Japan}
\affiliation{RIKEN Nishina Center, Wako 351-0198, Japan}

\author{Jing Peng}
\email{jpeng@bnu.edu.cn}
\affiliation{School of Physics and Astronomy, Beijing Normal University, Beijing 100875, China}
\affiliation{Key Laboratory of Multiscale Spin Physics (Ministry of Education), Beijing Normal University, Beijing 100875, China}

\author{Haozhao Liang}
\email{haozhao.liang@phys.s.u-tokyo.ac.jp}
\affiliation{Department of Physics, Graduate School of Science, The University of Tokyo, Tokyo 113-0033, Japan}
\affiliation{Quark Nuclear Science Institute, The University of Tokyo, Tokyo 113-0033, Japan}
\affiliation{RIKEN Center for Interdisciplinary Theoretical and Mathematical Sciences (iTHEMS), Wako 351-0198, Japan}

\date{\today}

\begin{abstract}
We present the first quantitative microscopic realization of topologically quantized alignment in a finite nuclear system. The realization is obtained by exact diagonalization of a cranking seniority model, with the first Chern number evaluated over the sphere of cranking-axis orientations and analyzed together with the orientation-averaged alignment and cranking-frame configuration probabilities. The Chern number changes in integer steps as the system evolves from initially paired configurations to increasingly aligned configurations. A new intermediate phase is found in which the Chern number is already nonzero while the alignment continues to evolve toward its quantized value. We show that this deviation originates from the competition among pairing, axial quadrupole splitting, and Coriolis mixing. Thus, our microscopic approach reveals a more nuanced emergence of topologically quantized alignment in realistic nuclei, providing a quantitative stepping stone toward experimental investigations.
\end{abstract}

\maketitle

\emph{Introduction.}---Topological invariants characterize the global structure of quantum states over parameter manifolds and remain unchanged under smooth deformations of the system, provided that the relevant energy gap does not close~\cite{Nakahara2003,Berry1984,Simon1983}. Such a manifold is in general specified by the parameters on which a Hamiltonian depends and over which its eigenstates are followed. Adiabatic transport around closed paths on this manifold gives geometric phases known as Berry phases~\cite{Berry1984,Simon1983}. The integral of the associated Berry curvature over a closed two-dimensional manifold yields the first Chern number~\cite{Nakahara2003}. These concepts underlie a broad range of topological phenomena, from the integer quantum Hall effect~\cite{TKNN1982,Niu1985} and electronic topological insulators~\cite{HasanKane2010} to Chern band physics realized with ultracold fermions in the Haldane model~\cite{Jotzu2014}. Recently, the realization of a Laughlin state with two rapidly rotating ultracold fermions further demonstrates that rotation can provide a route to quantum Hall physics in a finite interacting system~\cite{Lunt2024}. Within nuclear structure, Berry phases have been investigated in rotating pair transfer, backbending, and seniority-conserving spectra~\cite{Nikam1987,Ring2013,ValienteDobon2021}. Rapidly rotating nuclei offer a natural setting for applying this geometric viewpoint to rotational alignment, with the cranking-axis orientation sphere serving as the relevant parameter manifold.

Rotational alignment in rapidly rotating nuclei is most naturally described in the rotating frame. Pairing correlations favor time-reversed partners and oppose pair breaking, while the Coriolis term lowers configurations aligned with the rotational axis and drives quasiparticle alignment~\cite{Mottelson1960,Stephens1972,Stephens1975,Frauendorf2001}. Backbending and upbending patterns inferred from level spacings identify the frequency region where alignment sets in Refs.~\cite{Stephens1975,Frauendorf2001,Capponi2020}. Aligned angular momenta extracted relative to a rotating core provide a direct diagnostic of quasiparticle alignment and band crossings. Particle rotor, cranking shell model, cranking Nilsson--Strutinsky and total Routhian surface, rotating mean-field, and shell model configuration interaction approaches provide standard theoretical descriptions of aligned configurations, shape evolution, and band crossings~\cite{BohrMottelson1975,Bengtsson1979,Bengtsson1985,Nazarewicz1985,Ring1980,FrauendorfMeng1997,Frauendorf2001,Caurier2005}. However, a quantitative description of rotational alignment in terms of Berry curvature and Chern number on the cranking-axis orientation sphere remains largely unexplored.

Recently, Guidry and Sun~\cite{Guidry2026} introduced topologically quantized alignment (TQA) in rapidly rotating nuclei. They showed that the first Chern number $C_1$ associated with the cranking-axis orientation sphere quantizes a Hilbert-space averaged rotation-aligned angular momentum. An ideal TQA relation is written as $\langle\!\langle\hat J_n\rangle\!\rangle= |C_1|/2$, where  $\langle\!\langle\hat J_n\rangle\!\rangle$ is the orientation average of the angular momentum component $\hat J_n\equiv\hat{\bm J}\cdot\bm n$ along the cranking axis $\bm{n}$. Their construction separates three regions: a low-frequency paired region with negligible alignment and $C_1=0$, a conventional rotational alignment region in which alignment develops while $C_1$ remains zero, and a topological alignment region with $C_1\neq0$ satisfying the ideal TQA relation. However, TQA has not yet been demonstrated with a microscopic nuclear model or observed experimentally.

In this Letter, we provide the first quantitative realization of TQA. We propose an exactly diagonalized two-orbital seniority Hamiltonian that retains a minimal but essential set of high-spin ingredients: $J=0$ pairing, axial quadrupole splitting, and Coriolis mixing. The lowest eigenvectors are computed over the full orientation sphere and used consistently to evaluate $C_1$, the averaged alignment, and the cranking-frame configuration probabilities. Recovering the three-region structure proposed by Guidry and Sun~\cite{Guidry2026} in the exact diagonalization, we further identify an intermediate topological phase in which $C_1$ is already nonzero while the orientation-averaged alignment continues toward the ideal TQA relation.

\emph{Model.}---We consider a cranking seniority Hamiltonian for two identical nucleons occupying the spherical $1d_{3/2}$ and $1f_{7/2}$ orbitals across the $sd$--$pf$ shell gap near $N, Z=20$. This two-orbital space provides a minimal setting in which pair breaking, cross-shell promotion, and fully aligned configurations can be tracked within the same exact diagonalization. The Hamiltonian depends on the cranking frequency $\omega$ and cranking-axis orientation $(\beta,\gamma)$ and reads
\begin{align}
    \hat H(\beta,\gamma;\omega)
    =& \sum_{\alpha m}\epsilon_\alpha
    \hat{a}^\dagger_{\alpha m}\hat{a}_{\alpha m}
    +\sum_{\alpha m}\chi_\alpha m^2
    \hat{a}^\dagger_{\alpha m}\hat{a}_{\alpha m}
    \nonumber\\
    &-\sum_{\alpha\alpha'}G_{\alpha\alpha'}
    \hat S_\alpha^\dagger \hat S_{\alpha'}
    -\omega\,\hat{\bm J}\cdot\bm n(\beta,\gamma),
    \label{eq:H}
\end{align}
where $\alpha=d,f$ labels the two spherical orbitals with $j_d=3/2$ and $j_f=7/2$ and $m=-j_\alpha,\ldots,j_\alpha$ denotes the projection on the intrinsic deformation axis $J_3$. The first term gives the spherical single-particle energies, with $\epsilon_f-\epsilon_d=\Delta E$ defining the $d_{3/2}$--$f_{7/2}$ energy separation. The second term is the one-body axial quadrupole splitting associated with deformation. It resolves the four $d_{3/2}$ and eight $f_{7/2}$ magnetic substates, and in our convention positive $\chi_\alpha$ raises high-$|m|$ substates. The third term is the $J=0$ pairing interaction between time-reversed partners. The pair operator is
\begin{equation}
    \hat S_\alpha^\dagger
    =
    \sum_{m>0}(-1)^{j_\alpha-m}\hat{a}^\dagger_{\alpha, m}\hat{a}^\dagger_{\alpha,-m}.
\end{equation}
The coefficients $G_{\alpha\alpha'}$ specify pair-scattering strengths within and between the two orbitals. The off-diagonal terms $G_{df}=G_{fd}\ne0$ allow a $J=0$ pair to scatter between the $d_{3/2}$ and $f_{7/2}$ orbitals. The last term in Eq.~(\ref{eq:H}) is the Coriolis term, which couples the total angular momentum to the cranking axis $\bm n(\beta,\gamma)$. The projected angular momentum is
\begin{equation}
    \hat J_n(\beta,\gamma)
    =\hat{\bm J}\cdot\bm n
    =\hat J_1\sin\beta\cos\gamma+\hat J_2\sin\beta\sin\gamma+\hat J_3\cos\beta,
    \label{eq:jn_projection}
\end{equation}
where $\beta$ is measured from $J_3$ and $\gamma$ is measured in the $J_1$--$J_2$ plane.
\begin{figure}[tbp]
    \centering
    \includegraphics[width=\columnwidth]{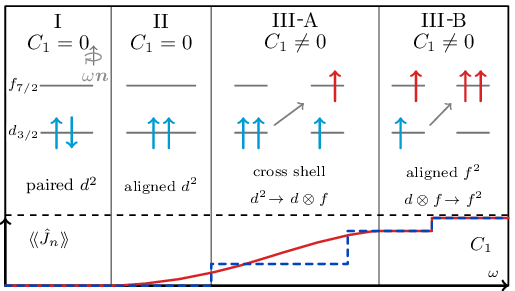}
    \caption{Schematic configuration evolution considered in this work. Increasing cranking frequency drives the system from paired $d_{3/2}^2$ to aligned $d_{3/2}^2$, then to $d_{3/2}\otimes f_{7/2}$, and finally to aligned $f_{7/2}^2$. The lower panel shows schematic $\langle\!\langle \hat J_n\rangle\!\rangle$ (solid line) and $C_1$ (dashed line) versus the rotational frequency $\omega$.}
    \label{fig:mechanism_cartoon}
\end{figure}

The Hamiltonian~(\ref{eq:H}) is exactly diagonalized for each $(\beta,\gamma;\omega)$ in the antisymmetrized two-particle Hilbert space spanned by the $4+8=12$ magnetic substates, whose dimension is $\binom{12}{2}=66$. The $d_{3/2}$--$f_{7/2}$ single-particle spacing is chosen to represent the normal $sd$--$pf$ shell gap around the doubly magic $^{40}$Ca core. Spectroscopy of the neighboring $A=39$ and $A=41$ nuclei identifies the relevant $d_{3/2}$ hole and $f_{7/2}$ particle states~\cite{Chen2018NDS39,Nesaraja2016NDS41}, and shell-model analyses of $^{40}$Ca provide a consistent reference scale~\cite{Caurier2007}. We therefore take $\epsilon_d=0$ and $\epsilon_f=\Delta E=7.0$~MeV. The axial quadrupole strength $\chi$ is chosen to produce a weak deformation splitting of the magnetic substates. Guided by Nilsson single-particle systematics and constrained covariant density functional theory spectra, we use the representative value $\chi_d=\chi_f=0.10$~MeV~\cite{Nilsson1955,BohrMottelson1975,Zhao2010PCPK1}. The seniority-pairing strength $G$ is guided by empirical odd-even mass staggering in this mass region, for which we take $G=0.50$~MeV~\cite{Satula1998}. Following the standard seniority-pairing convention \cite{Ring1980}, all pair-scattering coefficients are assigned the common value $G_{\alpha\alpha'}=G$. The cranking frequency $\omega$ is varied as a theoretical control parameter for tracing the Chern-number steps and the corresponding orientation-averaged alignment response.

Figure~\ref{fig:mechanism_cartoon} summarizes the configuration evolution driven by the competing terms in Eq.~(\ref{eq:H}).
Since the two orbitals have opposite parity, the schematic should be understood as the dominant configuration in the lowest eigenstate family of the cranking Hamiltonian at each $\omega$. An experimentally observed rotational band would also have to satisfy the usual parity and electromagnetic transition selection rules. At low frequency, the $J=0$ pairing term dominates, maintaining a time-reversed paired state that suppresses alignment. As $\omega$ increases, the Coriolis term lowers configurations with large projection along $\bm n$ and promotes pair breaking. This produces the idealized sequence shown in Fig.~\ref{fig:mechanism_cartoon} from the paired state to an aligned $d_{3/2}^2$ configuration, then to a $d_{3/2}\otimes f_{7/2}$ configuration, and finally to the aligned $f_{7/2}^2$ limit. The lower curves are qualitative and distinguish the continuous alignment response from the discrete changes of $C_1$. Quantitative phase assignment requires the exact eigenstates over the full orientation sphere, from which both $C_1$ and $\langle\!\langle \hat J_n\rangle\!\rangle$ are computed.
\begin{figure*}[t]
    \centering
    \includegraphics[width=0.80\textwidth]{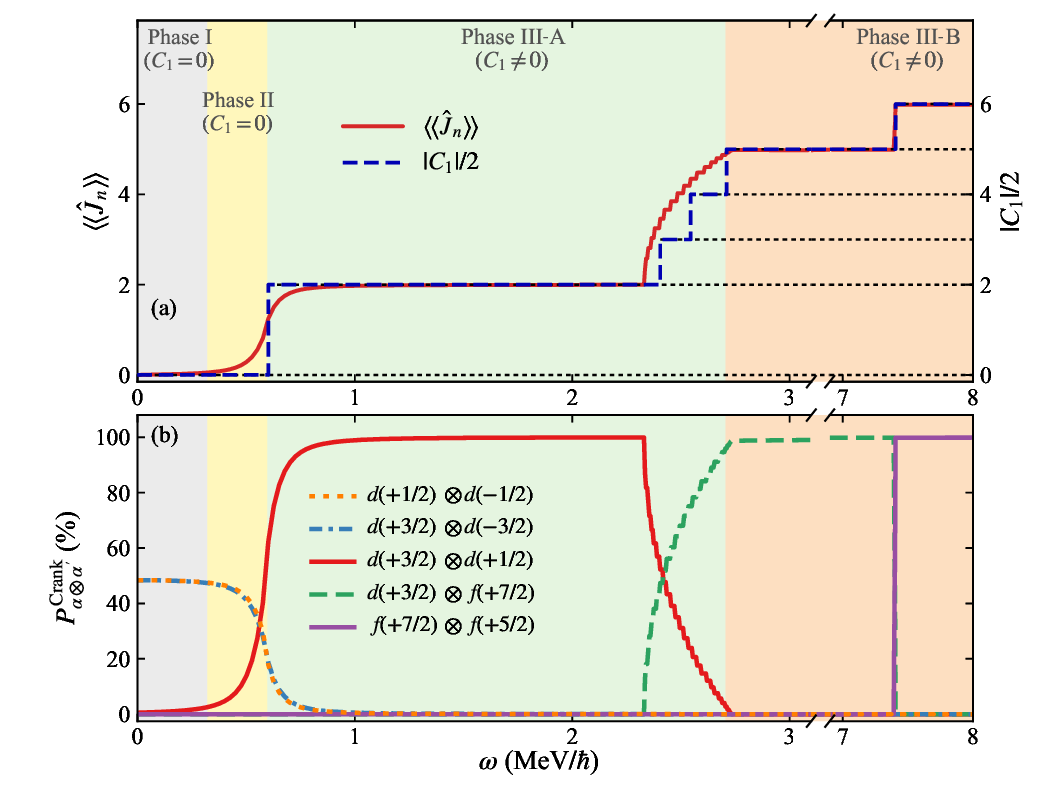}
    \caption[Stepwise changes of the Chern number in the exact two-nucleon space.]{Stepwise changes of the Chern number in the exact two-nucleon $1d_{3/2}\oplus1f_{7/2}$ model space for $\Delta E=7.0$~MeV, $G=0.50$~MeV, and $\chi_f=\chi_d=0.10$~MeV. (a) The orientation average $\langle\!\langle \hat J_n\rangle\!\rangle$ versus the quantity $|C_1|/2$. (b) Probabilities $P_{\alpha\otimes\alpha'}^{\rm Crank}$ of the dominant two-particle configurations in the cranking frame. The magnetic labels in (b) are projections along the instantaneous cranking axis $\bm n$. The interval $3.1<\omega<6.9$~MeV$/\hbar$ is omitted for display.}
    \label{fig:topological_sequence}
\end{figure*}

To connect this rotating-frame configuration evolution with topology, we construct the wave function over the full sphere of cranking-axis orientations. Let $E_k(\beta,\gamma;\omega)$ and $|\psi_k(\beta,\gamma;\omega)\rangle$ denote the eigenvalues and eigenvectors of $\hat H(\beta,\gamma;\omega)$ where $k$ labels the eigenstates in order of increasing energy at each $(\beta,\gamma;\omega)$. At fixed $\omega$, the Chern number is well defined when the lowest eigenstates remain gapped from the first excited states over the full orientation sphere, i.e., $E_1(\beta,\gamma;\omega)-E_0(\beta,\gamma;\omega)>0$ for all $(\beta,\gamma)$. The corresponding Chern number, Berry curvature, and Berry connection are
\begin{align}
    C_1(\omega)
    &=
    \frac{1}{2\pi}\int_{S^2} d\beta\,d\gamma\,F_{\beta\gamma},\nonumber\\
    F_{\beta\gamma}
    &=
    \partial_\beta A_\gamma-\partial_\gamma A_\beta,
    \qquad
    A_\mu=i\braket{\psi_0|\partial_\mu\psi_0},
    \label{eq:chern_continuum}
\end{align}
respectively.
The ideal TQA relation then reads
\begin{equation}
    \langle\!\langle \hat J_n\rangle\!\rangle= |C_1|/2 ,
    \label{eq:tqa_relation}
\end{equation}
with the averaged alignment defined as
\begin{equation}
    \langle\!\langle \hat J_n\rangle\!\rangle
    =
    \frac{1}{4\pi}\int d\Omega\,\bra{\psi_0(\beta,\gamma;\omega)}\hat{\bm J}\cdot\bm n\ket{\psi_0(\beta,\gamma;\omega)}.
    \label{eq:alignment_average}
\end{equation}
Since $\partial_\omega \hat H=-\hat J_n$, the Feynman--Hellmann theorem~\cite{Feynman1939} gives 
$
    \partial_\omega E_0(\beta,\gamma;\omega)
    = -\langle \psi_0(\beta,\gamma;\omega)|\hat J_n|\psi_0(\beta,\gamma;\omega)\rangle
    \equiv
    -\langle \hat J_n\rangle_{\beta\gamma}.
    \label{eq:fh_alignment}
$ 
The orientation-averaged alignment is an expectation value and can evolve continuously with $\omega$. The Chern number is an integer topological invariant that remains unchanged as long as the gap $E_1-E_0$ stays nonzero everywhere on $S^2$ and can change only through a gap closing at some orientation. At such a frequency, the wave function is degenerate at that orientation, so no Chern number is assigned. For each $\omega$, $C_1$ is evaluated on the orientation mesh with the gauge-invariant lattice algorithm of Fukui, Hatsugai, and Suzuki~\cite{Fukui2005}, and $\langle\!\langle \hat J_n\rangle\!\rangle$ in Eq.~(\ref{eq:alignment_average}) is computed with the $\sin\beta$ quadrature weight. To identify the dominant cranking-frame configurations, the same eigenvectors are projected onto two-particle configurations $\ket{\Phi_{\alpha\otimes\alpha'}(\bm n)}$ quantized along the instantaneous cranking axis, where $\alpha\otimes\alpha'$ labels the two occupied orbitals in that frame. The corresponding cranking-frame probability is
\begin{equation}
    P_{\alpha\otimes\alpha'}^{\rm Crank}(\omega)
    =
    \frac{1}{4\pi}\int d\Omega\,
    \left|\braket{\Phi_{\alpha\otimes\alpha'}(\bm n)|\psi_0(\beta,\gamma;\omega)}\right|^2 ,
    \label{eq:crank_probability}
\end{equation}
with $d\Omega=\sin\beta\,d\beta\,d\gamma$. Further details are given in the Supplemental Material~\cite{SM}.

\emph{TQA from exact diagonalization.}---Figure~\ref{fig:topological_sequence} presents two complementary diagnostics obtained from the same exact eigenvectors. Panel (a) compares the discrete Chern number $|C_1|/2$ with $\langle\!\langle \hat J_n\rangle\!\rangle$. Panel (b) resolves the corresponding wave functions into cranking-frame configurations.

As the cranking frequency increases, the Chern number changes in integer steps, giving
$
|C_1|/2=0\rightarrow2\rightarrow3\rightarrow4\rightarrow5\rightarrow6.
$
The first nonzero step is associated with the aligned $d_{3/2}^2$ configuration where the Pauli exclusion principle forbids the $d(+1/2)\otimes d(+1/2)$ configuration and the cranking term favors $d(+3/2)\otimes d(+1/2)$, causing $|C_1|/2$ to jump directly to $2$; the intermediate steps accompany the evolution toward $d_{3/2}\otimes f_{7/2}$, and the final value corresponds to the fully aligned $f_{7/2}^2$. $\langle\!\langle \hat J_n\rangle\!\rangle$ evolves continuously across these steps and approaches the corresponding $|C_1|/2$ values over finite frequency intervals. We quantify the deviation from the ideal TQA relation by
\begin{equation}
    \delta_J(\omega)=|\langle\!\langle \hat J_n\rangle\!\rangle-|C_1|/2|
\end{equation}
and take $\delta_J=0.05$ as the numerical tolerance for satisfying Eq.~(\ref{eq:tqa_relation}). The schematic picture of Guidry and Sun~\cite{Guidry2026} contains two $C_1=0$ regions, Phases I and II, and one ideal topological region, Phase III, with $C_1\neq0$. The exact calculation resolves this $C_1\neq0$ region into Phase III-A with $\delta_J>0.05$ and Phase III-B with $\delta_J<0.05$.

With low cranking frequency, the system is in Phase I, with $C_1=0$ and $\delta_J<0.05$.
The corresponding $k=0$ eigenstates are paired $d_{3/2}^2$ state mainly composed of the two time-reversed configurations $d(+3/2)\otimes d(-3/2)$ and $d(+1/2)\otimes d(-1/2)$. The topology is trivial, and the orientation-averaged alignment is negligible within the numerical tolerance.

Phase II is a topologically trivial alignment region with $C_1=0$ but finite orientation-averaged alignment. The finite alignment is generated by the Coriolis mixing in the two-particle Hilbert space. The probabilities in the cranking frame show that the aligned configuration $d(+3/2)\otimes d(+1/2)$ is dominant before $C_1$ becomes nonzero.

The newly identified Phase III-A is an intermediate topological region with $C_1\neq0$ in which agreement with the ideal TQA relation is not sustained as $\omega$ increases. The microscopic origin is visible in Fig.~\ref{fig:topological_sequence}(b). The dominant component first becomes aligned $d_{3/2}^2$ and then changes to $d(+3/2)\otimes f(+7/2)$ around $\omega\simeq2.4$--$2.8$~MeV$/\hbar$. In high-spin terminology, this is the two-orbital analogue of a band crossing in the rotating frame. The dominant configuration changes when Coriolis energy gain compensates the costs of pair breaking and promotion across the $sd$--$pf$ shell gap. The probability remains concentrated in a few configurations. Their evolution reflects the competition among $J=0$ pairing, axial quadrupole splitting, and Coriolis alignment.

Phase III-B is the nontrivial topological region ($C_1\neq0$) in which the ideal TQA remains satisfied within
the numerical tolerance as $\omega$ is further increased. The Phase III-A--III-B boundary lies close to a nearby Chern-number step. The lower-frequency part of Phase III-B satisfies $\langle\!\langle\hat J_n\rangle\!\rangle\simeq |C_1|/2=5$ and is dominated by the $d(+3/2)\otimes f(+7/2)$ configuration. A further Chern step near $\omega\simeq7.4$ MeV$/\hbar$ gives the final value $|C_1|/2=6$, where the fully aligned $f(+7/2)\otimes f(+5/2)$ configuration carries about $99.9\%$ of the cranking-frame probability. In short, the Chern number changes by integer steps and the averaged alignment reaches the corresponding $|C_1|/2$ value only after the cranking-frame configuration becomes essentially pure and aligned.

\begin{figure}[tbp]
    \centering
    \includegraphics[width=\columnwidth]{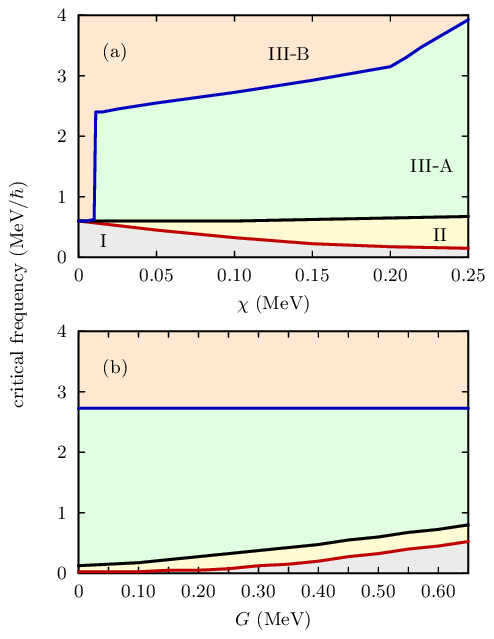}
    \caption[Critical frequencies]{Critical frequencies as functions of the axial quadrupole splitting $\chi$ and the pairing strength $G$. (a) Dependence on $\chi_f=\chi_d=\chi$ with a fixed $G=0.50$~MeV. (b) Dependence on $G$ with fixed $\chi_f=\chi_d=0.10$~MeV. The curves denote the Phase I--II, II--III-A, and III-A--III-B boundaries; shaded regions follow the conventions in Fig.~\ref{fig:topological_sequence}.}
    \label{fig:phase_boundaries}
\end{figure}

\emph{Critical frequencies.}---The phase boundaries quantify how pairing, axial splitting, and Coriolis mixing control the emergence of TQA. Figure~\ref{fig:phase_boundaries} shows the dependence of the corresponding critical frequencies on $G$ and $\chi$. The critical frequencies are defined numerically as the minimum values of $\omega$ satisfying
\begin{equation}
\begin{aligned}
\omega_{\mathrm{I|II}}
&=
\min\Big\{\omega \,:
\langle\!\langle\hat J_n\rangle\!\rangle \ge 0.05
\Big\},\\
\omega_{\mathrm{II|III-A}}
&=
\min\Big\{\omega \,:
|C_1|/2>0
\Big\},\\
\omega_{\mathrm{III-A|III-B}}
&=
\min\Big\{\omega \,:
\delta_J(\omega')\le0.05\ 
\forall\,\omega'\ge\omega
\Big\}.
\end{aligned}
\end{equation}
All three boundaries are extracted from the exact eigenstates, Chern numbers, and averaged alignments computed on the same $\omega$ grid as in Fig.~\ref{fig:topological_sequence}.

Figure~\ref{fig:phase_boundaries}(a) shows that, at $\chi=0$, the axial quadrupole term vanishes and the Coriolis term selects aligned configurations directly. The alignment crossover then coincides with the onset of nonzero $C_1$, giving the ideal TQA behavior in this model. For $\chi>0$, the axial quadrupole term is diagonal in the intrinsic $J_3$ basis and the cranking term acts along $\bm n$. These two terms generally do not commute over $S^2$, allowing $C_1$ to become nonzero before the averaged alignment reaches the corresponding quantized value.

For very small $\chi$, the first interval with $C_1\neq0$ already satisfies $\delta_J<0.05$, so the exact solution proceeds directly from Phase II into Phase III-B. With increasing $\chi$, a finite III-A interval opens during the $d_{3/2}^2\to d_{3/2}\otimes f_{7/2}$ evolution, where $C_1\neq0$ and $\delta_J>0.05$. The sharp initial rise of the III-A--III-B boundary occurs when the first interval with $C_1\neq0$ no longer satisfies the alignment criterion, so recovery of Eq.~(\ref{eq:tqa_relation}) is delayed until the later $|C_1|/2=5$ interval. For moderate $\chi$, this boundary tracks the frequency region where the dominant configuration changes from $d_{3/2}^2$ to $d_{3/2}\otimes f_{7/2}$ for orientations $\bm n$ close to $J_3$. At larger $\chi$, it bends upward because the exact eigenstate must suppress the residual mismatch generated by the noncommuting axial quadrupole and Coriolis terms before the alignment criterion is recovered. Thus, axial quadrupole splitting separates the observable angular momentum response from the underlying topological transition.

Figure~\ref{fig:phase_boundaries}(b) shows that the Phase I--II and II--III-A boundaries increase monotonically with increasing $G$, while the change of the Phase III-A--III-B boundary is negligible. This behavior shows that pairing mainly controls the low-frequency onset of alignment and nonzero $C_1$. Finite pairing stabilizes the time-reversed $d_{3/2}^2$ pair against Coriolis breaking and delays the onset of the phase with $C_1 \neq 0$. Once the wave function is aligned, the recovery of the relation $\langle\!\langle\hat J_n\rangle\!\rangle= |C_1|/2$ is governed primarily by the shell gap and the axial quadrupole splitting, insensitive to the pairing strength.

\emph{Spectroscopic implications.}---$C_1$ and $\langle\!\langle\hat J_n\rangle\!\rangle$ are defined in the cranking Hamiltonian, providing a theoretical link between topology and rotational alignment. Their spectroscopic relevance can be assessed through level energies, spin assignments, electromagnetic transition patterns, and aligned angular momenta extracted from measured rotational bands. Such aligned angular momenta are conventionally defined relative to a rotating core, for example with the Harris prescription~\cite{Harris1965}. Backbending or upbending systematics then identify the frequency interval where quasiparticle alignment or a band crossing develops~\cite{Stephens1975,Frauendorf2001,Capponi2020}. In the present mechanism, such anomalies would signal the alignment response and the accompanying change of intrinsic configuration. Previous nuclear Berry-phase studies show that geometric phases can leave signatures in reaction or spectroscopic observables, for example in rotating pair transfer and in seniority-conserving spectra~\cite{Nikam1987,Ring2013,ValienteDobon2021}. A realistic search for TQA should therefore look for a characteristic evolution of spin alignment together with independent evidence for a change in intrinsic configuration, constrained where possible by electromagnetic or transfer matrix elements. In addition, Guidry and Sun~\cite{Guidry2026} suggested anomalous stability of high-spin states against gamma decay or fission as a possible global signature of TQA. The present reduced Hamiltonian establishes the connection between topological quantization and rotational alignment at the eigenstate level. Future extensions to a dynamical framework are needed to connect this relation to experimental observables.

\emph{Summary.}---We have provided the first quantitative realization of TQA in an interacting nuclear model, using an exactly diagonalized two-orbital cranking seniority Hamiltonian. When the lowest eigenstates remain isolated on the cranking-axis orientation sphere, the first Chern number is well-defined and changes in integer steps as the cranking frequency is varied. The same eigenvectors give orientation-averaged alignment and the cranking-frame configuration content, enabling a direct comparison of topology, alignment, and microscopic configuration evolution.

The exact diagonalization further identifies an intermediate topological region, Phase III-A, in which $C_1\neq0$ and the averaged alignment continues toward the value $|C_1|/2$. This intermediate region arises from the competition among $J=0$ pairing, axial quadrupole splitting, and Coriolis mixing. We further introduce a quantitative phase diagram that distinguishes Phases I, II, III-A, and III-B and determines the corresponding critical frequencies. The phase boundaries exhibit distinct dependences on pairing and axial quadrupole splitting. Pairing stabilizes the low-frequency paired state, and the axial quadrupole splitting separates the onset of nonzero $C_1$ from the realization of the ideal TQA. The relation $\langle\!\langle\hat J_n\rangle\!\rangle = |C_1|/2$ is recovered when the eigenstate is dominated by a single aligned configuration in the cranking-frame basis. Thus, the exact solution establishes a microscopic realization of the quantized topological structure proposed for rapidly rotating nuclei and provides a quantitative basis for future experimental searches for TQA. Our work further motivates extending topologically quantized alignment to larger rotating systems, where richer integer phases and possible fractional counterparts may emerge.

\begin{acknowledgments}
\emph{Acknowledgments.}---The authors are grateful to Wenmin Deng for assistance with the numerical calculations. S.W. acknowledges funding from the China Scholarship Council (CSC) (File No.~[202406050068]), the National Natural Science Foundation of China under grant No.~12575130, and the Chongqing Natural Science Foundation under grant No.~CSTB2025NSCQ-GPX0742. 
This work is also supported by the JSPS KAKENHI Grant Numbers JP24H00239, JP26K07063, and JP26K01431. 
\end{acknowledgments}

\bibliography{refe}

\end{document}


\raggedbottom

\begin{center}
{\large\bfseries Supplemental Material: Microscopic Realization of Topologically Quantized Alignment in Fast-Rotating Nuclei}\\[0.4em]
Ganlong Ding, Sibo Wang, Hiroyuki Tajima, Daisuke Suzuki, Jing Peng, and Haozhao Liang
\end{center}
\vspace{1em}

This Supplemental Material presents the model parameters, the numerical details of the Chern number, the derivation of the ideal TQA relation, the definition of the cranking-frame configuration probabilities, the evolution of $\langle\!\langle\hat J_n\rangle\!\rangle$ and $C_1$ over the complete frequency interval, and the dependence of the phase boundaries on $G$ and $\chi$. Unless stated otherwise, numbered sections, equations, figures, and tables refer to this Supplemental Material. The term ``the paper'' refers to the accompanying paper ``Microscopic Realization of Topologically Quantized Alignment in Fast-Rotating Nuclei.''

\section{Model Hamiltonian and Parameters}
The model Hamiltonian and notation used below are the same as in Eqs.~(1) and (2) of the paper. The $m^2$ dependence of the axial splitting term follows from the diagonal matrix elements of an axial quadrupole one-body operator. By the Wigner--Eckart theorem \cite{Edmonds1960}, these matrix elements satisfy
\begin{equation}
    \langle j_\alpha m|\hat Q_{20}|j_\alpha m\rangle
    =Q_\alpha\frac{3m^2-j_\alpha(j_\alpha+1)}{j_\alpha(2j_\alpha-1)},
    \qquad
    Q_\alpha
    =\langle j_\alpha j_\alpha|\hat Q_{20}|j_\alpha j_\alpha\rangle .
    \label{eq:supp_quadrupole_m}
\end{equation}
The contribution $-j_\alpha(j_\alpha+1)$ is independent of $m$ within a fixed orbital and is included in $\epsilon_\alpha$. The remaining magnetic-substate dependence is proportional to $m^2$ and is represented by $\chi_\alpha m^2$. Thus, $\Delta E$ specifies the spherical $d_{3/2}$--$f_{7/2}$ separation, and the $\chi_\alpha$ contains the reduced quadrupole matrix element and the strength of the axial deformation \cite{Nilsson1955,BohrMottelson1975}.

The pair operators are defined in Eq.~(2) of the paper. The normalized $J=0$ pair states are
\begin{equation}
    |\alpha^2;J=0\rangle
    =\frac{1}{\sqrt{\Omega_\alpha}}\,\hat S_\alpha^\dagger|0\rangle,
    \qquad
    \Omega_\alpha=\frac{2j_\alpha+1}{2},
\end{equation}
where $\Omega_\alpha$ is the number of time-reversed pairs in orbital $\alpha$. For the pairing term in Eq.~(1) of the paper, the matrix elements between these states are
\begin{equation}
\langle \alpha^2;J=0|\hat H_{\rm pair}|\alpha'^2;J=0\rangle = -G_{\alpha\alpha'}\sqrt{\Omega_\alpha\Omega_{\alpha'}} .
\end{equation}
Thus, setting $G_{\alpha\alpha'}=G$ assigns a common coefficient to all pair-scattering channels, while the corresponding matrix elements retain their orbital dependence through $\Omega_d=2$ and $\Omega_f=4$.

\begin{figure}[htbp]
    \centering
    \includegraphics[width=0.50\linewidth]{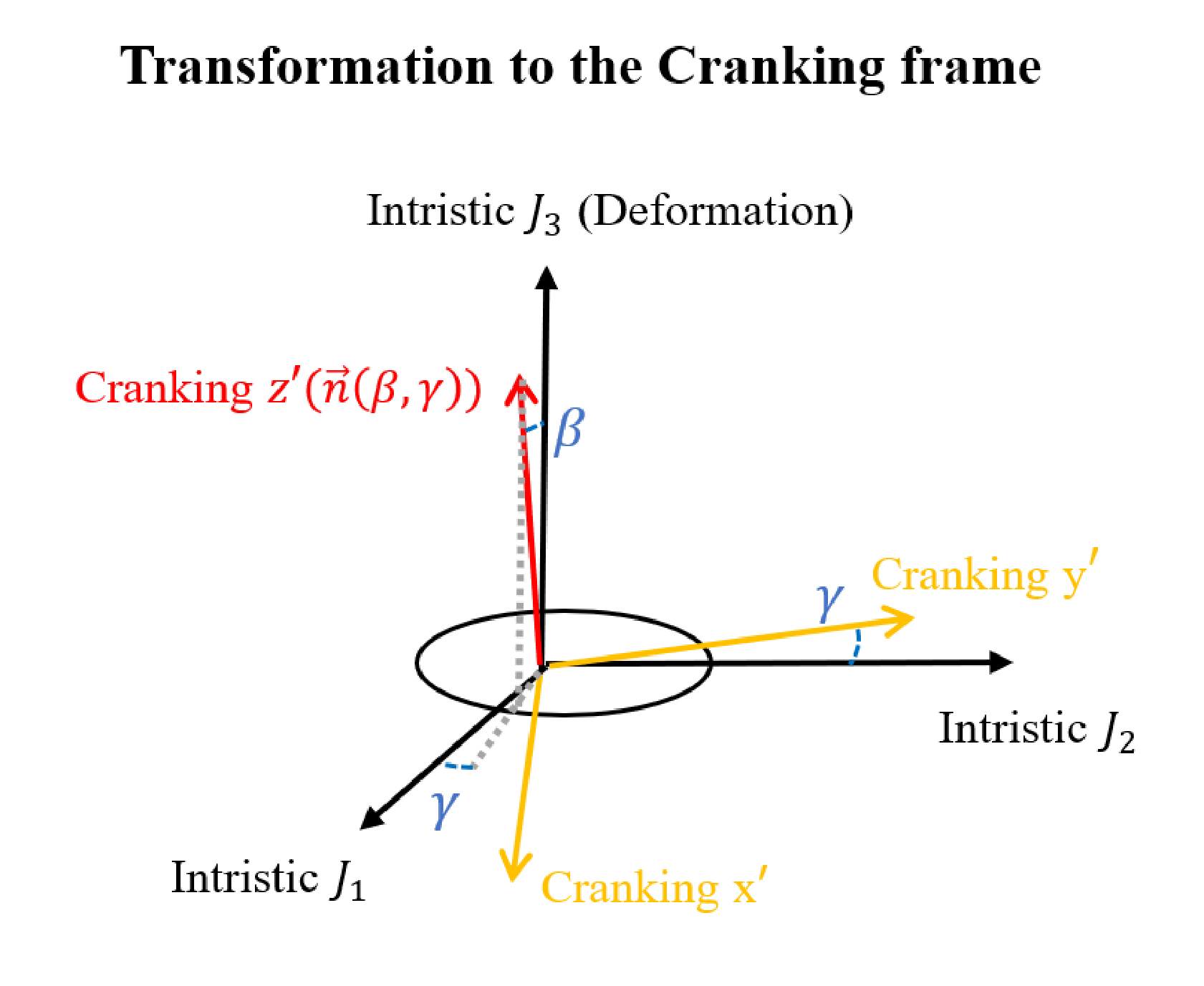}
    \caption{Intrinsic axes and cranking-axis orientation. The cranking frequency vector is $\bm\omega=\omega\bm n(\beta,\gamma)$, where $\beta$ is measured from the intrinsic $J_3$ axis and $\gamma$ is measured in the intrinsic $J_1$--$J_2$ plane.}
    \label{fig:supp_frame}
\end{figure}

The Coriolis term introduces the dependence on the orientation of the cranking axis. Figure~\ref{fig:supp_frame} defines the corresponding rotational geometry. The angle $\beta$ is measured from the intrinsic $J_3$ axis, and $\gamma$ is measured from the intrinsic $J_1$ axis in the $J_1$--$J_2$ plane. The two angles specify the orientation of the cranking axis and are distinct from the nuclear quadrupole deformation variables. We define
\begin{equation}
\bm n(\beta,\gamma)=(\sin\beta\cos\gamma,\sin\beta\sin\gamma,\cos\beta),
\end{equation} 
with the corresponding angular momentum component
\begin{equation}
\hat J_n(\beta,\gamma)=\hat J_1\sin\beta\cos\gamma+\hat J_2\sin\beta\sin\gamma+\hat J_3\cos\beta .
\label{eq:supp_Jn}
\end{equation}
Including the Coriolis term, the Hamiltonian becomes
\begin{equation}
\hat H(\beta,\gamma;\omega)=\hat H_0-\omega \hat J_n(\beta,\gamma).
\label{eq:supp_H_orientation}
\end{equation}

\section{Numerical Evaluation of the Chern Number}
At each fixed $\omega$, exact diagonalization provides the eigenvalues $E_k(\beta_i,\gamma_j;\omega)$ and eigenvectors $|\psi_k(\beta_i,\gamma_j;\omega)\rangle$ on the cranking-axis orientation mesh, where $k$ labels the eigenstates in order of increasing energy. We evaluate the first Chern number using the gauge-invariant lattice method of Fukui, Hatsugai, and Suzuki \cite{Fukui2005}. Let $\ell=(i,j)$ denote the mesh point $(\beta_i,\gamma_j)$. The displacements $\mathbf e_1$ and $\mathbf e_2$ advance the mesh by one point in the $\beta$ and $\gamma$ directions, respectively, so that $\ell+\mathbf e_1=(i+1,j)$ and $\ell+\mathbf e_2=(i,j+1)$. The mesh is periodic in $\gamma$, with $\gamma=0$ and $2\pi$ identified. At $\beta=0$ and $\pi$, all values of $\gamma$ represent the same physical orientation. In the following equations, $\mu=1,2$ labels the two mesh directions. The link variables are
\begin{equation}
U_\mu(\ell)
=\frac{\braket{\psi_0(\ell)|\psi_0(\ell+\mathbf e_\mu)}}{\left|\braket{\psi_0(\ell)|\psi_0(\ell+\mathbf e_\mu)}\right|}.
\end{equation}
The lattice field strength on each elementary plaquette is
\begin{equation}
F_{12}(\ell)=\operatorname{Arg}\left[U_1(\ell)U_2(\ell+\mathbf e_1)U_1^{-1}(\ell+\mathbf e_2)U_2^{-1}(\ell)\right].
\end{equation}
Here, $\operatorname{Arg}$ is taken on its principal branch $(-\pi,\pi]$. The lattice Chern number is
\begin{equation}
C_1(\omega)=\frac{1}{2\pi}\sum_\ell F_{12}(\ell).
\end{equation}
The Chern number is defined only when the lowest eigenstate family remains nondegenerate over the full orientation sphere. This condition is expressed as
\begin{equation}
\Delta_{\min}(\omega)=\min_{\beta,\gamma}\left[E_1(\beta,\gamma;\omega)-E_0(\beta,\gamma;\omega)\right]>0 .
\label{eq:supp_global_gap}
\end{equation}
When Eq.~(\ref{eq:supp_global_gap}) is not satisfied, $\psi_0$ is degenerate at one or more orientations and $C_1$ is undefined. In addition, to resolve the closely spaced low-frequency changes in $C_1$ and the high-frequency transition, different meshes and frequency steps are used in the corresponding intervals. Unless otherwise specified, the calculations use $\Delta\omega=0.025$ MeV$/\hbar$, reduced to $0.005$MeV$/\hbar$ for $2.25\leq\omega\leq2.90$ MeV$/\hbar$ and to $0.002$ MeV$/\hbar$ for $7.34\leq\omega\leq7.46$ MeV$/\hbar$. The orientation meshes and frequency ranges are summarized in Table~\ref{tab:supp_numerics}.
\begin{table}[H]
    \caption{Numerical details for figures in the paper. Frequencies are given in MeV$/\hbar$.}
    \label{tab:supp_numerics}
    \begin{ruledtabular}
    \begin{tabular}{lll}
    Figure & $(\beta,\gamma)$ mesh & $\omega$ range \\
    \hline
    Fig.~2(a)
    & $21\times21$
    & $\omega\in[0,2.25)\cup(2.90,10]$ \\
    Fig.~2(a)
    & $41\times41$
    & $\omega\in[2.25,2.90]$  \\
    Fig.~2(b)
    & $61\times91$
    & $\omega\in[0,10]$ \\
    Fig.~3
    & $21\times21$
    & $\omega\in[0,10]$
    \end{tabular}
    \end{ruledtabular}
\end{table}

\section{Ideal TQA Relation for a Single Aligned Configuration}
The ideal TQA relation follows when the $k=0$ eigenstate at each orientation corresponds to a single aligned configuration with a definite angular-momentum projection along the instantaneous cranking axis. Angular momenta are expressed in units of $\hbar$, as in the paper. Consider a two-particle configuration $\alpha m_\alpha\otimes\alpha' m_{\alpha'}$, where $m_\alpha$ and $m_{\alpha'}$ are the single-particle projections along $\bm n(\beta,\gamma)$. In the ideal aligned limit,
\begin{equation}
    \hat J_n|\psi_0(\beta,\gamma)\rangle=(m_\alpha+m_{\alpha'})|\psi_0(\beta,\gamma)\rangle .
    \label{eq:supp_ideal_Jn}
\end{equation}
A convenient gauge for the state at a general orientation is obtained by rotating the north-pole state,
\begin{equation}
    |\psi_0(\beta,\gamma)\rangle
    =e^{-i\gamma\hat J_3}e^{-i\beta\hat J_2}|\psi_0(0,0)\rangle ,\qquad\hat J_3|\psi_0(0,0)\rangle
    =(m_\alpha+m_{\alpha'})|\psi_0(0,0)\rangle .
    \label{eq:supp_ideal_state}
\end{equation}
For the Berry connection $A_\mu=i\braket{\psi_0|\partial_\mu\psi_0}$, with $\mu=\beta,\gamma$, Eq.~(\ref{eq:supp_ideal_state}) gives
\begin{align}
    A_\beta &=0, \nonumber\\
    A_\gamma
    &=\bra{\psi_0(0,0)}e^{i\beta\hat J_2}\hat J_3e^{-i\beta\hat J_2}\ket{\psi_0(0,0)}\nonumber =(m_\alpha+m_{\alpha'})\cos\beta .
    \label{eq:supp_ideal_connection}
\end{align}
Here, we have used
\begin{equation}
    e^{i\beta\hat J_2}\hat J_3e^{-i\beta\hat J_2}=\hat J_3\cos\beta-\hat J_1\sin\beta
\end{equation}
and
$\langle\psi_0(0,0)|\hat J_1|\psi_0(0,0)\rangle=0$
for a state with a definite $J_3$ projection. The Berry curvature is then
\begin{equation}
    F_{\beta\gamma}=\partial_\beta A_\gamma-\partial_\gamma A_\beta=-(m_\alpha+m_{\alpha'})\sin\beta .
    \label{eq:supp_ideal_curvature}
\end{equation}
With the orientation convention adopted here, the corresponding Chern number is
\begin{equation}
    C_1=\frac{1}{2\pi}\int_0^{2\pi}d\gamma\int_0^\pi d\beta\,
    F_{\beta\gamma}=-2(m_\alpha+m_{\alpha'}) .
    \label{eq:supp_ideal_chern}
\end{equation}
Its magnitude therefore satisfies
\begin{equation}
    |C_1|=2|m_\alpha+m_{\alpha'}|.
\end{equation}
For the aligned configurations considered here, $m_\alpha+m_{\alpha'}>0$. Combining this result with Eq.~(\ref{eq:supp_ideal_Jn}) gives
\begin{equation}
    \langle\!\langle\hat J_n\rangle\!\rangle=m_\alpha+m_{\alpha'}=\frac{|C_1|}{2}.
    \label{eq:supp_ideal_tqa}
\end{equation}
The double brackets in Eq.~(\ref{eq:supp_ideal_tqa}) denote the $4\pi$-normalized orientation average defined in the paper. Equation~(16) of Guidry and Sun \cite{Guidry2026} instead relates the Chern number to the Hilbert-space average of the rotation-aligned angular-momentum component $\hat J_2$ in their axially symmetric quadrupole-rotor model. The two averages are defined over different spaces and consequently carry different normalization factors. Equation~(\ref{eq:supp_ideal_tqa}) is the corresponding relation for a fixed two-particle aligned configuration on the cranking-axis orientation sphere used here. The configurations $d(+3/2)\otimes d(+1/2)$, $d(+3/2)\otimes f(+7/2)$, and $f(+7/2)\otimes f(+5/2)$ therefore correspond to $|C_1|/2=2$, $5$, and $6$, respectively. Deviations from Eq.~(\ref{eq:supp_ideal_tqa}) quantify the departure of the system from a single aligned configuration in the cranking frame.

\section{Configuration Probabilities in the Cranking Frame}
\begin{figure}[htbp]
    \centering
    \includegraphics[width=0.70\linewidth]{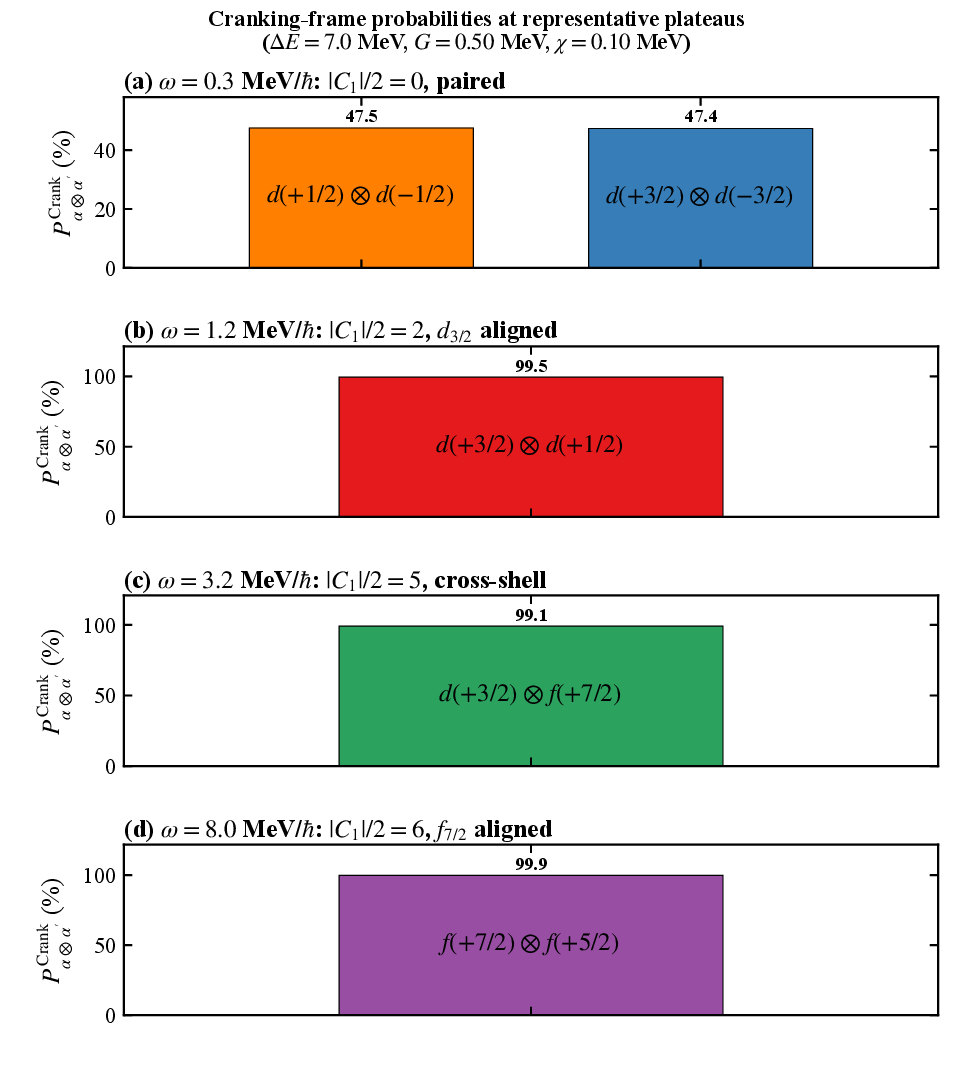}
    \caption{Cranking-frame configuration probabilities at selected frequencies for the parameter set used in the paper. The magnetic labels denote projections along the instantaneous cranking axis.}
    \label{fig:supp_config_bars}
\end{figure}
At each point on the cranking-axis orientation sphere, the lowest eigenvector is expanded in the fixed intrinsic two-particle basis as
\begin{equation}
    |\psi_0(\beta,\gamma;\omega)\rangle
    =\sum_{\alpha\otimes\alpha'}C_{\alpha\otimes\alpha'}(\beta,\gamma;\omega)|\Phi_{\alpha\otimes\alpha'}\rangle .
    \label{eq:supp_local_expansion}
\end{equation}
Here $|\Phi_{\alpha\otimes\alpha'}\rangle$ is an antisymmetrized two-particle Slater determinant quantized along the intrinsic $J_3$ axis. The orientation-averaged probability in the fixed intrinsic basis is
\begin{equation}
    P_{\alpha\otimes\alpha'}^{\rm Int}(\omega)=\frac{1}{4\pi}\int d\Omega\,|C_{\alpha\otimes\alpha'}(\beta,\gamma;\omega)|^2,
    \qquad
    d\Omega=\sin\beta\,d\beta\,d\gamma .
    \label{eq:supp_P_int}
\end{equation}
A configuration with definite projections along the instantaneous cranking axis is represented by different superpositions of intrinsic magnetic substates as the cranking axis moves over $S^2$. The probabilities used in the paper are therefore evaluated in the instantaneous cranking frame. The rotation from the intrinsic $J_3$ axis to $\bm n(\beta,\gamma)$ is generated by
\begin{equation}
    U(\beta,\gamma)=e^{-i\gamma\hat J_3}e^{-i\beta\hat J_2},
    \qquad
    U(\beta,\gamma)\hat J_3U^\dagger(\beta,\gamma)=\hat J_n(\beta,\gamma).
    \label{eq:supp_rotation}
\end{equation}
The transformation first tilts the intrinsic $J_3$ axis by $\beta$ about the intrinsic $J_2$ axis and then rotates the tilted axis by $\gamma$ about the original intrinsic $J_3$ axis. This sequence produces the direction $\bm n(\beta,\gamma)$ defined in Eq.~(\ref{eq:supp_Jn}) and also fixes the transverse $x'$ and $y'$ axes shown in Fig.~\ref{fig:supp_frame}. The adjoint rotation is
\begin{equation}
    U^\dagger(\beta,\gamma)=e^{i\beta\hat J_2}e^{i\gamma\hat J_3}.
    \label{eq:supp_rotation_dagger}
\end{equation}
For a two-particle determinant $|\Phi_{\alpha\otimes\alpha'}\rangle$ quantized along the intrinsic $J_3$ axis, the corresponding determinant quantized along $\bm n(\beta,\gamma)$ is
\begin{equation}
    |\Phi_{\alpha\otimes\alpha'}^{(n)}(\beta,\gamma)\rangle
    =U(\beta,\gamma)|\Phi_{\alpha\otimes\alpha'}\rangle .
    \label{eq:supp_rotated_determinant_intro}
\end{equation}
The corresponding cranking-frame amplitude is
\begin{equation}
    \widetilde C_{\alpha\otimes\alpha'}(\beta,\gamma;\omega)
    =\langle\Phi_{\alpha\otimes\alpha'}|U^\dagger(\beta,\gamma)|\psi_0(\beta,\gamma;\omega)\rangle .
    \label{eq:supp_Ctilde}
\end{equation}
The orientation-averaged cranking-frame probability is then
\begin{equation}
    P_{\alpha\otimes\alpha'}^{\rm Crank}(\omega)=
    \frac{1}{4\pi}\int d\Omega\,|\widetilde C_{\alpha\otimes\alpha'}(\beta,\gamma;\omega)|^2 .
    \label{eq:supp_P_crank}
\end{equation}
Equation~(\ref{eq:supp_P_crank}) projects the eigenvector at each orientation onto a configuration quantized along the instantaneous cranking axis and averages the resulting probability over the orientation sphere.

On the numerical mesh, Eqs.~(\ref{eq:supp_P_int}) and (\ref{eq:supp_P_crank}) are evaluated by quadrature. For uniform grids in $\beta$ and $\gamma$,
\begin{equation}
    P_{\alpha\otimes\alpha'}^{\rm Crank}(\omega)
    \simeq
    \frac{\sum_{ij} w_i\left|\sum_{\bar\alpha\otimes\bar\alpha'}R_{\alpha\otimes\alpha',\bar\alpha\otimes\bar\alpha'}(\beta_i,\gamma_j)C_{\bar\alpha\otimes\bar\alpha'}(\beta_i,\gamma_j;\omega)\right|^2}{\sum_{ij}w_i},
    \qquad
    w_i=\sin\beta_i .
    \label{eq:supp_P_crank_mesh}
\end{equation}
The normalization $\sum_{\alpha\otimes\alpha'}P_{\alpha\otimes\alpha'}^{\rm Crank}=1$ is satisfied. The probabilities in Fig.~\ref{fig:supp_config_bars} refer to two-particle determinants quantized along the instantaneous cranking axis. For example, $P_{d(+3/2)\otimes f(+7/2)}^{\rm Crank}$ is the orientation-averaged probability of finding one $d_{3/2}$ particle and one $f_{7/2}$ particle with the indicated projections along that axis. For $\chi=0.10$ MeV, the probability of the fully aligned $f(+7/2)\otimes f(+5/2)$ configuration approaches unity at high frequency. The residual admixture arises because the axial quadrupole term is diagonal in the intrinsic $J_3$ basis, whereas the Coriolis term and the cranking-frame projection are defined along $\bm n$. This mixing vanishes at $\chi=0$, where the same orientation mesh and projection yield, at $\omega=10.0$ MeV$/\hbar$,
\begin{equation}
    1-P^{\rm Crank}_{f(+7/2)\otimes f(+5/2)}
    =3.7\times10^{-15},
\end{equation}
showing that the remaining deviation from unity is at the level of numerical roundoff.
\begin{figure}[htbp]
    \centering
    \includegraphics[width=0.85\linewidth]{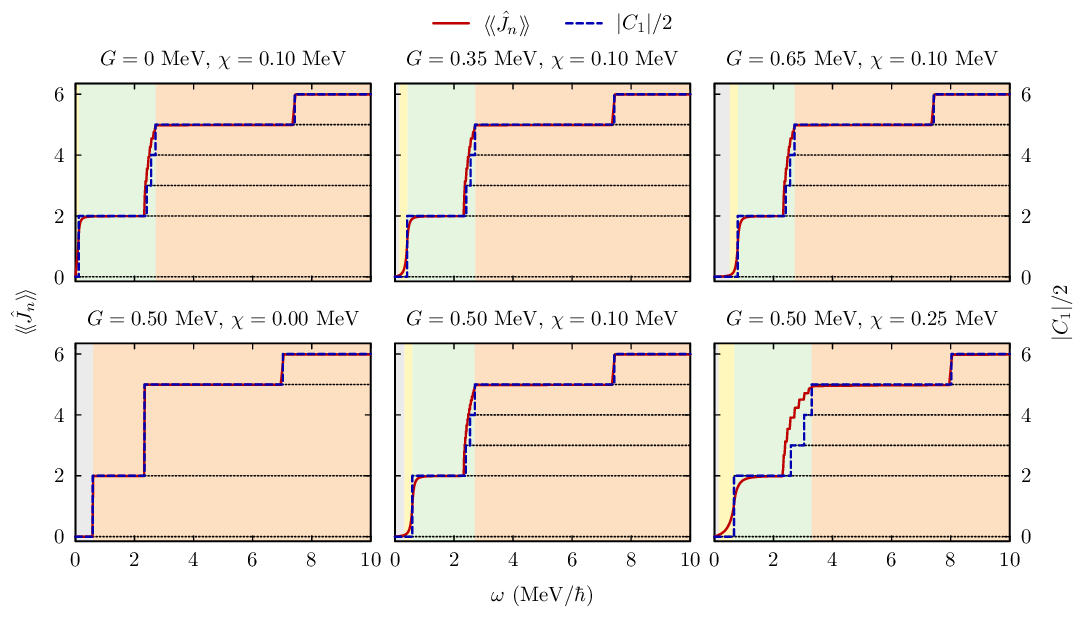}
    \caption{Evolutions of $\langle\!\langle\hat J_n\rangle\!\rangle$ and $|C_1|/2$ as funcitons of $\omega$ with different pairing strength $G$ and axial quadrupole splitting $\chi$. The upper row varies $G$ at fixed $\chi=0.10$ MeV, and the lower row varies $\chi$ at fixed $G=0.50$ MeV. The phases are assigned using the definitions in the paper.}
    \label{fig:supp_parameter_variants}
\end{figure}

\section{Additional Numerical Results}
Figure~\ref{fig:supp_parameter_variants} shows the evolutions of $\langle\!\langle\hat J_n\rangle\!\rangle$ and $C_1$ with different $G$ and $\chi$. At $G=0$, the pairing term vanishes, but the integer steps remain. Increasing $G$ shifts the Phase I--II and II--III-A boundaries to higher frequencies, with a smaller effect on the Phase III-A--III-B boundary. In contrast, increasing $\chi$ enhances the deviation between the continuous alignment response and the quantized values of $|C_1|/2$, thereby enlarging Phase III-A and delaying the emergence of the ideal TQA relation. At $\chi=0$, the magnetic substates remain degenerate with respect to the axial quadrupole term, and the agreement between $\langle\!\langle\hat J_n\rangle\!\rangle$ and $|C_1|/2$ is closest to the ideal TQA relation. At high frequency, all parameter sets approach the same fully aligned limit, $\langle\!\langle\hat J_n\rangle\!\rangle = |C_1|/2 =6$.

\bibliography{refe}